\documentclass[prb,superscriptaddress,twocolumn,floatfix]{revtex4-1}

\usepackage[FIGTOPCAP]{subfigure}
\usepackage[usenames]{color}
\usepackage{amsmath,amssymb}
\usepackage[pdftex]{hyperref,graphicx}
\hypersetup{colorlinks = true, urlcolor = blue, linkcolor = blue, citecolor = blue}

\begin{document}
\title{Topological superconductivity, ferromagnetism, and valley-polarized phases in moir\'e systems: Renormalization group analysis for twisted double bilayer graphene}
\author{Yi-Ting~Hsu}
\email{ythsu@umd.edu}
\affiliation{Condensed Matter Theory Center and Joint Quantum Institute, University of Maryland, College Park, MD 20742, USA}
\author{Fengcheng Wu}
\affiliation{Condensed Matter Theory Center and Joint Quantum Institute, University of Maryland, College Park, MD 20742, USA}
\author{S. Das Sarma}
\affiliation{Condensed Matter Theory Center and Joint Quantum Institute, University of Maryland, College Park, MD 20742, USA}
\date{\today}

\begin{abstract}
Recent experiments have observed possible spin- and valley-polarized insulators and spin-triplet superconductivity in twisted double bilayer graphene, a moir\'e structure consisting of a pair of Bernal-stacked bilayer graphene. 
Besides the continuously tunable band widths controlled by an applied displacement field and twist angle, these moir\'e bands also possess van Hove singularities near the Fermi surface and a field-dependent nesting which is far from perfect.  
Here we carry out a perturbative renormalization group analysis to unbiasedly study the competition among all possible instabilities in twisted double bilayer graphene and related systems with a similar van Hove fermiology in the presence of weak but finite repulsive interactions. 
Our key finding is that there are several competing magnetic, valley, charge, and superconducting instabilities arising from interactions in twisted double bilayer graphene, which can be tuned by controlling the displacement field and the twist angle. 
In particular, we show that spin- or valley-polarized uniform instabilities generically dominate under moderate interactions smaller than the band width, whereas $p$-wave spin-triplet topological superconductivity and exotic spin-singlet modulated paired state become important as the interactions decrease. 
Realization of our findings in general moir\'e systems with a similar van Hove fermiology should open up new opportunities for manipulating topological superconductivity and spin- or valley-polarized states in highly tunable platforms. 
\end{abstract}

\maketitle
\section{Introduction}
Moir\'e systems, which comprise two atomically thin Van der Waals layers with a relative twist angle, have recently attracted extensive attention for their highly controllable band structure and many-body interactions, leading to interesting physics in regimes from weak- to strong-coupling 
\cite{Exp_SC_BLG_Cao2018,Exp_CI_BLG_Cao18,Exp_DBLG_Kim,Exp_TriGraphene_CI_FengWang,Exp_MoireWSe2}. 
In particular, a rich variety of interaction-driven phases have been discovered in various moir\'e systems 
that are controllable using the twist angle, external displacement field, and electric gating
\cite{Exp_SC_BLG_Cao2018,Exp_CI_BLG_Cao18,Exp_BLG_sc_Dean,Exp_BLG_nematic,Exp_BLG_nematic_nphys,
Exp_FM_BLG,Exp_BLG_StrangeMetal,Exp_BLG_LinearT_Young,Exp_BLGsmallangle_CI,Exp_BLG_Chern_Barcelona,Exp_BLG_compressibility_PRL,
Exp_DBLG_Kim,Exp_SC_DBLG_Zhang19,Exp_DBLG_Pablo19,Exp_VPCI_DBLG_Tutuc19,Exp_TriGraphene_CI_FengWang,Exp_TriGraphene_FMChernI_FengWang,Exp_MoireWSe2}. 
For instance, correlated insulating states emerge in commensurately-filled twisted bilayer graphene\cite{Exp_CI_BLG_Cao18,Exp_BLG_Chern_Barcelona,Exp_BLG_compressibility_PRL,Exp_BLGsmallangle_CI,Exp_BLG_nematic} with signatures of ferromagnetism in certain cases\cite{Exp_FM_BLG,Exp_BLG_Chern_Barcelona,serlin2020intrinsic}, and superconductivity appear over a wider range of carrier density\cite{Exp_SC_BLG_Cao2018,Exp_BLG_Chern_Barcelona,Exp_BLG_sc_Dean,Exp_BLGsmallangle_CI}. 
More recently, experimental evidences for both spin-\cite{Exp_DBLG_Kim,Exp_SC_DBLG_Zhang19,Exp_DBLG_Pablo19} and valley-polarized\cite{Exp_VPCI_DBLG_Tutuc19} insulating phases as well as correlated metallic phases with spontaneous symmetry breaking\cite{Exp_DBLG_Cmetallic} were found in twisted double bilayer graphene (TDBG) under an external displacement field at various fillings. 
Signatures of spin-triplet superconductivity, of which critical temperature increases with increasing in-plane magnetic fields, have also been reported in TDBG \cite{Exp_DBLG_Kim}.
Given the rich possibilities of symmetry-broken phases and the multi-dimensional parameter space waiting to be explored, it is desirable to have theoretical frameworks that extract essential features out of the complicated microscopic models and identify the stable phases when scanning through experimentally relevant parameters. The problem is subtle and difficult because of the large number of symmetry-allowed phases and phase transitions possibly competing in flat band moire systems at low twist angles\cite{RG_honeycomb_Robert}.

While many prior theoretical studies on moir\'e systems focus on strong coupling approaches\cite{Toposc_XuBalents,BLG_StrongU_SC_stanford,
Thy_BLG_MottSc_Senthil_PRX,Thy_BLG_StrongU_Vafek,Thy_DBLG_ncomm} due to their nearly flat bands enhancing interaction effects at low twist angles\cite{Bistritzer2011}, weak-coupling approaches were also adapted to interpret the observed phases as arising from various Fermi surface instabilities\cite{KLsc_BLG_Stauber_PRL,PatchRG_nonesting_PRB,RPA_BLG_chiraldsc_YangPRL,
RPA_BLG_valleyflucSc_You,
RG_Moire_Fu_PRX,RG_Moire_Nandkishore,Wu2018phonon,wu2019phonon,wu2018topological,wu2019identification,Thy_DBLGpair_Scheurer} with the justification that the measured interaction-driven energy gaps are typically smaller than the band width\cite{Exp_SC_BLG_Cao2018, Exp_CI_BLG_Cao18,Exp_BLG_compressibility_PRL}. 
Although the competition among the various instabilities is known to be sensitive to the details of the Fermi surface and the underlying moire band structure, the existence of van Hove singularities near the Fermi level, which is a common feature shared among the moir\'e bands \cite{Exp_BLG_nematic,BLGmodel_VHS_Fu_PRX}, allows considerable simplifications of the problem. 
Since the density of states diverges (at least) logarithmically near the van Hove (VH) points and presumably govern the main physics, instead of treating the full Fermi surface, one could simplify the problem by keeping only patches centered at the VH points in the instability analysis. 
Under such a VH patch approximation with patch sizes much smaller than the moir\'e Brillouin zone, a perturbative renormalization group (RG) technique dubbed parquet RG\cite{Schulz_RG_1987,Salmhofer_RG_1998,RG_FeSc_Chubukov_PRB,RG_FeSc_Fernandes_PRX,
RG_oddparitysc_Yao} has been applied to monolayer\cite{Nphys_Nandkishore2012} and twisted bilayer graphene\cite{PatchRG_nonesting_PRB,RG_Moire_Fu_PRX,RG_Moire_Nandkishore} to study how inter- and intra-patch interactions can lead to dominant instabilities. 
Nonetheless, perturbative RG studies of such kind often show a strong preference towards density waves and even-parity superconductivity (in the absence of symmetries that enforce degeneracy between even- and odd-parity superconductivity)  \cite{RG_FeSc_Chubukov_PRB,Nphys_Nandkishore2012,RG_Moire_Fu_PRX,RG_Moire_Nandkishore}, even for systems away from perfect nesting\cite{RG_Moire_Fu_PRX,RG_Moire_Nandkishore}. 
The propensity of the patch RG theory to lead to density wave and even-parity superconductivity, which is also found within the simplest mean field theories, arises from the effective one dimensional nature of the 'patch system' where the VH points act as the 1D Fermi points in the nested 2D twisted material. 
Such a framework therefore seems even qualitatively incapable to describe moir\'e systems that are plausible candidates for uniform symmetry-broken phases and spin-triplet superconductivity, such as TDBG, since, e.g., density wave instabilities do not seem to dominate the low energy physics of the experimental moire narrow band systems. 

In the following, we explain how this weak-pairing approach, but not necessarily within the simplest patch approximation, can in fact serve as a general theoretical framework treating systems with VH points, including those that are prone to spin-triplet superconductivity and uniform symmetry-broken phases. 
For a given instability, its tendency for becoming dominant can be quantified by the product $V(E)\Pi(\textbf{q},E)$ between its driving interaction $V(E)$ and the associated bare susceptibility $\Pi(\textbf{q},E)$ at momentum \textbf{q}. 
Studying the competition among the instabilities then amounts to identifying the instability with the largest tendency as the energy scale $E$ decreases towards the Fermi surface. 

In the infinitesimal interaction limit, it is well known that superconductivity in general wins over particle-hole instabilities since the particle-particle susceptibility $\Pi_{pp}(0,E)\sim$ln$^2(\Lambda/E)$ diverges as log square when $E\rightarrow 0$, whereas the particle-hole susceptibility $\Pi_{ph}(\textbf{q},E)\sim$ln$(\Lambda/E)$ diverges at most logarithmically.  
The particle-hole susceptibility at some large momentum \textbf{Q} can only diverge as log square when the Fermi surface is perfectly nested. In such cases, the density waves modulated at \textbf{Q} competes with $d$-wave superconductivity, which is enhanced by corresponding fluctuations. Odd-parity superconductivity, on the other hand, could become competitive only when these two instabilities are suppressed by insufficient Fermi surface nesting and when the patches at opposite momenta are not related by reciprocal lattice vectors\cite{RG_oddparitysc_Yao}. 
   
When the interaction strength becomes finite but still much smaller than the band width, which is the case for most realistic weakly interacting systems, the driving interactions could diverge at a non-vanishing critical energy scale $E_c>0$, which thereby sets an early cutoff to the slow-growing difference between the ln and ln$^2$ functions. 
In this case, the associated bare susceptibility alone does not fully determine the competition outcome, and a uniform particle-hole instability 
could dominate as well if its driving interaction overcomes the difference in bare susceptibilities at $E_c$. 
Furthermore, the RG flows of these driving interactions also become parametrically sensitive to both the ln- and ln$^2$-growing contributions (instead of just the latter) 
due to the early cutoff set by $E_c$. 
It is therefore crucial to include both ln- and ln$^2$-growing contributions throughout the RG analysis in order to unbiasedly identify the dominant instability for weak-coupling systems away from the infinitesimal interaction limit, especially for those with at most moderately nested Fermi surface\cite{RG_oddparitysc_Yao}.
In this paper, we perform such an unbiased study in a complete fashion, which has not been done previously in moir\'e systems to our knowledge. 
We find that keeping the competition between the ln and ln$^2$ terms is important in determining the possible phases of the moire system at finite interactions.

\begin{figure}[t]
\includegraphics[width=8cm]{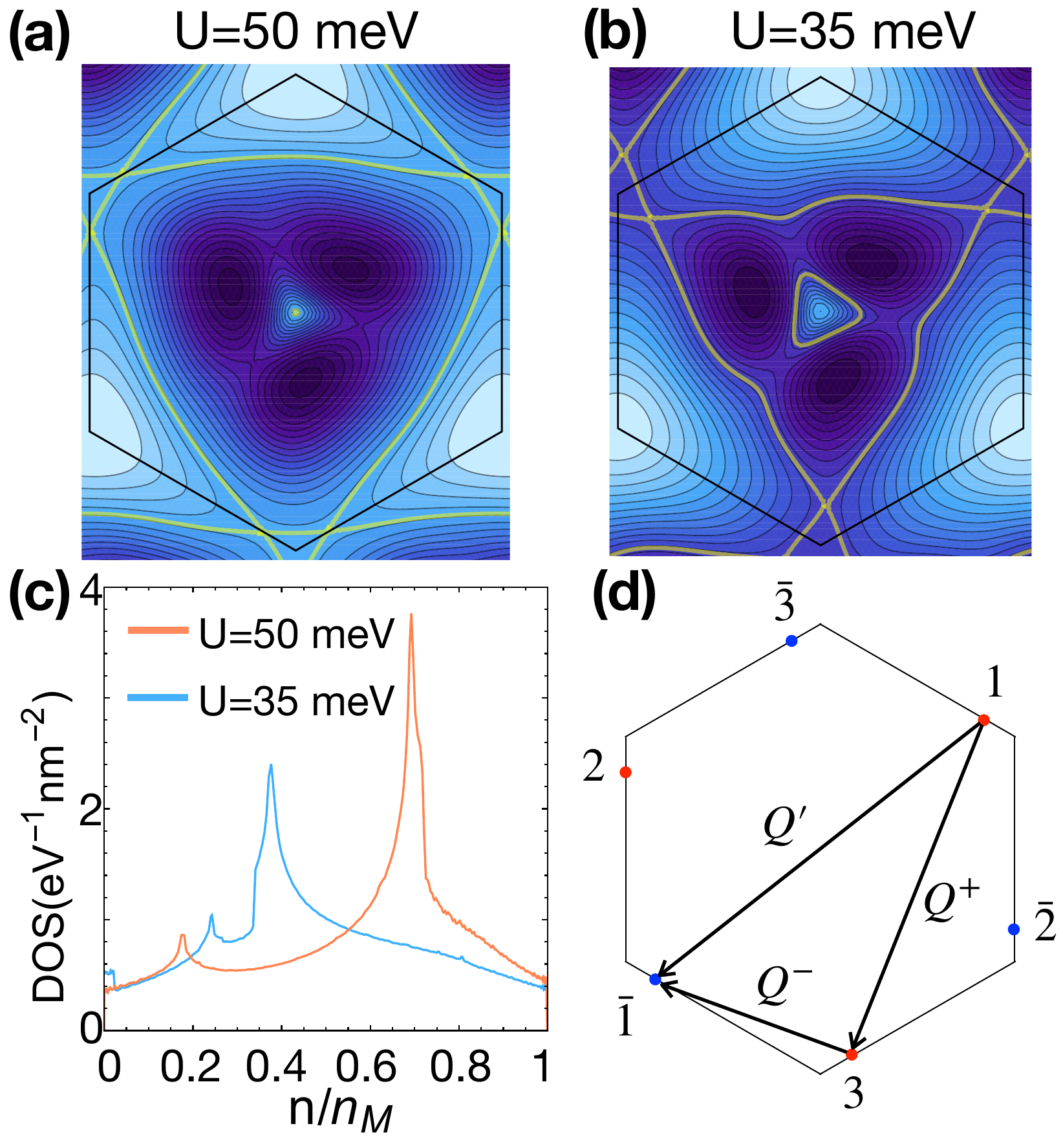}
\caption{(a) Energy contour plot for the first moir\'e conduction band
	in $+K$ valley of TDBG with a twist angle $\theta=1.24^{\circ}$ and a layer dependent potential $U=50$ meV that is generated by an out-of-plane displacement field. See Ref.~\onlinecite{wu2019Ferro} for details of the moir\'e Hamiltonian that leads to the band structure. (b) Similar as (a) but for $U=35$ meV. The yellow lines in (a) and (b) mark the Fermi surfaces at the van Hove energy, which are tunable by the displacement field. (c) The calculated density of states as a function of filling factor for the bands shown in (a) and (b). The VHS we consider correspond to the largest peak in each curve. 
(d) Schematics of the patch model we consider for the representative TDBG van Hove fermiology in (a). The hexagon, red points, and blue points represent the moir\'e Brillouin zone, the patch centers from $+K$ valley, and those from $-K$ valley, respectively. The arrows represent the characteristic momenta connecting the inter- and intra-valley patches. 
}
\label{FS}
\end{figure}
Our goal is to study the allowed weak-coupling phase diagram away from the infinitesimal interaction limit for TDBG and other moir\'e systems with a similar van Hove pattern, which consists of three van Hove singularities (VHS) per valley for two valleys related by time-reversal symmetry. 
Importantly, while the Fermi surface nesting degree is tunable by an external displacement field in TDBG [see Fig. \ref{FS}(a)(b) and section \ref{sec:model}], it is \textit{at most moderate} due to the lattice symmetry\cite{RG_Moire_Nandkishore}. Thus, theoretical results based on models assuming perfect (or close to perfect) nesting are inapplicable to TDBG. 
To unbiasedly treat all possible instabilities on equal footing, we adapt the perturbative parquet RG approach under patch approximation with all the ln-diverging contributions included throughout the analysis. 
Specifically, we study how the nine inequivalent intra- and inter-patch interactions arising from the considered van Hove fermiology evolve towards the long wavelength limit and lead to different dominant instabilities. This is different from the previous RG analysis\cite{PatchRG_nonesting_PRB}, where seven interactions were considered. 
As a result, we find that in the weakly nested regimes, spin-triplet topological superconductivity and a modulated paired state dominate in the weakly interacting limit, whereas spin-polarized and valley-polarized states appear for stronger interactions. 
Since both the interaction strength (relative to the band width) and the nesting degree are experimentally tunable via knobs 
such as the twist angle and external displacement field\cite{Thy_DBLG_ncomm,wu2019Ferro}, our results could offer useful guidance for future experimental exploration of exotic superconducting and metallic phases. In particular, our predicted topological superconductivity and the modulated phase should both be experimentally observable in the future.

The rest of the paper is structured as follows. In section \ref{sec:model}, we describe the non-interacting dispersions of TDBG, and show how the van Hove singularity patterns change under different displacement field strengths. In section \ref{sec:RGflow}, we show the RG calculation based on the non-interacting TDBG model step by step, including the key non-interacting susceptibilities, RG equations of the inter- and intra-patch interactions, and the tendencies for the considered instabilities. 
In section \ref{sec:PD}, we present the RG flows and the resulting phase diagrams in the absence and presence of inter-valley scatterings. Finally in section \ref{sec:summary}, we discuss the experimental relevance of our results.

\section{Van Hove fermiology in TDBG}\label{sec:model}
We use TDBG as a model system for the parquet RG study without assuming the 1D perfect nesting limit. TDBG consists of a pair of Bernal-stacked bilayer graphene twisted at a relative angle. 
Different from the twisted bilayer graphene of which lowest-energy bands are gapless and can only be nearly flat at fine-tuned magic twist angles, 
the first conduction band in TDBG can be energetically isolated from other bands  by applying a displacement field, and its band width can be further tuned by both the field and angle\cite{Thy_DBLG_ncomm,wu2019Ferro}. Thus, TDBG has more tunability as a moire system (both twist angle and displacement field) enabling, in principle, access to a richer quantum phase diagram than twisted bilayer graphene. 

To understand the general van Hove fermiology in TDBG, we examine its moir\'e band structure obtained from a microscopic model reported in Ref. \onlinecite{Thy_DBLG_ncomm}. 
In Fig. \ref{FS}(a)-(c), we show two representative moir\'e band structures of the first conduction band in $+K$ valley along with the corresponding density of states for two different displacement field strengths at certain twist angle. Here we note that the $+K$ valley originates from that of a Bernal-stacked bilayer graphene, and there is a counterpart $-K$ valley that is related to the $+K$ valley by spinless time-reversal symmetry. At the single particle level, moir\'e bands in $+K$ and $-K$ valleys can be studied separately.   
It is clear from both Fig.~\ref{FS}(a) and (b) that there are three inequivalent VHS related by threefold rotations per valley near the Fermi surface. 
The existence of these VHS allows us to apply the patch approximation, where we can focus only on patches centered at the three VH points \textbf{P}$_n$, $n=1,2,3$, with a patch size $k_\Lambda$ much smaller than the moir\'e Brillouin zone. 

Importantly, the positions and fillings at which these three VHS occur as well as the Fermi surface nesting degree within these patches can be tuned by the displacement field strength [see Fig. \ref{FS}(a)-(b)]. 
The low-energy dispersion within these three patches $n=1,2,3$ in $+K$ valley of TDBG can be described by the following general form  
\begin{align}
&\epsilon_{\textbf{k}}^1 \approx \sum_{\alpha=x,y} \sum_{\beta=x, y} w_{\alpha \beta} (\textbf{k}-\textbf{P}_1)_\alpha(\textbf{k}-\textbf{P}_1)_\beta\nonumber\\
&\epsilon_{\textbf{k}}^2 =  \epsilon_{\hat{\mathcal{R}}_3^{-1}\textbf{k}}^1, \,\,\, \epsilon_{\textbf{k}}^3 =  \epsilon_{\hat{\mathcal{R}}_3 \textbf{k}}^1,
\label{epsilon1}
\end{align}
where we keep only up to the quadratic terms in momentum $\textbf{k}$. Here, $\hat{\mathcal{R}}_3$ is the rotation matrix for $+2\pi/3$ rotation, and both $\textbf{k}$ and the VH point positions $\textbf{P}_1$, $\textbf{P}_2=\hat{\mathcal{R}}_3 \textbf{P}_1$, $\textbf{P}_3 = \hat{\mathcal{R}}_3^{-1} \textbf{P}_1$ are measured relative to the moir\'e Brillouin zone center $\bar{\Gamma}$ point. 
Both $\textbf{P}_n$ and the coefficient matrix $w$ are tunable by the displacement field. Here $w$ is a symmetric real matrix and obey  $\text{Det} (w) <0$ since $w$ describes dispersion around a saddle point. 
Since the Hamiltonians of the two valleys are related by time-reversal symmetry, there are three other VH patches $\bar{n}=\bar{1},\bar{2},\bar{3}$ from the $-K$ valley. The patch centers and the low-energy dispersions within the patches in the two valleys are related as $\textbf{P}_{\bar{n}}=-\textbf{P}_{n}$ and $\epsilon_{\textbf{k}}^{\bar{n}} = \epsilon_{-\textbf{k}}^{n}$. 
As an example, the patch dispersions for the case in Fig. \ref{FS}(a) are approximately given by 
\begin{align}
&\epsilon^1_{\textbf{k}}=-(k_y-P_{1y})^2-\sqrt{3}(k_y-P_{1y})(k_x-P_{1x})\nonumber\\
&\epsilon^2_{\textbf{k}}=\frac{1}{2}(k_y-P_{2y})^2-\frac{3}{2}(k_x-P_{2x})^2\nonumber\\
&\epsilon^3_{\textbf{k}}=-(k_y-P_{3y})^2+\sqrt{3}(k_y-P_{3y})(k_x-P_{3x})\nonumber\\
&\epsilon^{\bar{n}}_{\textbf{k}}=\epsilon^n_{-\textbf{k}}, ~~~n=1,2,3
\label{dispersion}
\end{align}
up to the quadratic terms in momentum $\textbf{k}$, where the patch centers are given by $\textbf{P}_i=(P_{ix},P_{iy})$. 

These six VH points $\textbf{P}_{n}$ and $\textbf{P}_{\bar{n}}$, $n=1,2,3$, are connected by vectors $\textbf{Q}'_n\equiv \textbf{P}_{\bar{n}}-\textbf{P}_{n}$, $\textbf{Q}^{+}_{nm}\equiv \textbf{P}_{m}-\textbf{P}_{n}$, and $\textbf{Q}^{-}_{n\bar{m}}\equiv \textbf{P}_{\bar{m}}-\textbf{P}_{n}$ [see Fig. \ref{FS}(d)], where patches $n$ and $m\neq n$ belong to the same valley. In terms of this notation, the dispersions of the opposite patches $n$ and $\bar{n}$ are related by $\epsilon^n_{\textbf{k}}=\epsilon^{\bar{n}}_{\textbf{k}+\textbf{Q}'_n}+O(k^3)$, which can be clearly seen in the example of Eq. (\ref{dispersion}).

\section{RG flows for the inter- and intra-patch interactions}\label{sec:RGflow}
\subsection{Bare susceptibilities}
The building blocks of the RG analysis are the intra- and inter-patch non-interacting static susceptibilities in the particle-hole and particle-particle channels 
\begin{align}
&\Pi_{\rm ph}^{nm}(\textbf{q})=-\int d\textbf{k}\frac{f_{\epsilon^n_{\textbf{k}}}
-f_{\epsilon^m_{\textbf{k}+\textbf{q}}}}{\epsilon^n_{\textbf{k}}-\epsilon^m_{\textbf{k}+\textbf{q}}}\nonumber\\
&\Pi_{\rm pp}^{nm}(\textbf{q})=\int d\textbf{k}\frac{1-f_{\epsilon^n_{\textbf{k}}}
-f_{\epsilon^m_{-\textbf{k}+\textbf{q}}}}{\epsilon^n_{\textbf{k}}+\epsilon^m_{-\textbf{k}+\textbf{q}}},  
\end{align}
where the patch indices $n$, $m=1,2,3,\bar{1},\bar{2},\bar{3}$. 
There are four important susceptibilities per channel at different momenta that connect associated patches. For the particle-hole channel, we have the density of states $\Pi_{\rm ph}^{nn}(0)$, the inter-valley susceptibilities with large momentum transfers $\Pi_{\rm ph}^{n\bar{n}}(\textbf{Q}'_n)$, and the susceptibilities manifesting the nesting degree for the Fermi surface (FS) of each valley $\Pi_{\rm ph}^{nm}(\textbf{Q}^+_{nm})$ and that between the FSs of the two valleys $\Pi_{\rm ph}^{n\bar{m}}(\textbf{Q}^-_{\bar{m}n})$, where $m\neq n$ belong to the same valley. For the particle-particle channel, we have the Cooper susceptibility $\Pi_{\rm pp}^{n\bar{n}}(0)$, the susceptibility 
for intra-patch pairing $\Pi_{\rm pp}^{nn}(-\textbf{Q}'_{n})$, and the susceptibilities for inter- and intra-valley nesting in the particle-particle channel $\Pi_{\rm pp}^{n\bar{m}}(\textbf{Q}^+_{\bar{n}\bar{m}})$ and $\Pi_{\rm pp}^{nm}(\textbf{Q}^-_{\bar{n}m})$ respectively. 

In particular, it is known\cite{Schulz_RG_1987,Salmhofer_RG_1998,VH_RG_Dzyaloshinskii,Nphys_Nandkishore2012,
PatchRG_nonesting_PRB,RG_Moire_Nandkishore} that the density of states and the Cooper instability exhibit ln and ln$^2$ divergences 
\begin{align}
&\Pi_{\rm ph}^{nn}(0)=\nu_0\ln\frac{\Lambda}{\text{max}(T,|\mu|)}\nonumber\\
&\Pi_{\rm pp}^{n\bar{n}}(0)=\frac{\nu_0}{2}\ln\frac{\Lambda}{\text{max}(T,|\mu|)}\ln\frac{\Lambda}{T}
\label{DOScooper}
\end{align}
due to the VHS, where the prefactor $\nu_0$ depends on the specific dispersions in Eq. \ref{epsilon1}, $\Lambda$ is the ultra-violate energy cutoff associated with the patch size $k_{\Lambda}$, $\mu$ is the chemical potential with respect to the VH points, and $T$ is the temperature.  
Importantly, in this work we focus on the realistic situation of being far from perfect nesting in both particle-hole and particle-particle channels. In such cases, the corresponding susceptibilities are ln- instead of ln$^2$-divergent, and we can parametrize them with respect to the density of states as  
\begin{align}
&\Pi_{\rm ph}^{n\bar{m}}(\textbf{Q}^-_{n\bar{m}})=\gamma_{3}\Pi_{\rm ph}^{nn}(0)\nonumber\\
&\Pi_{\rm pp}^{nm}(\textbf{Q}^-_{\bar{n}m})=\gamma_{5}\Pi_{\rm ph}^{nn}(0),  
\label{nesting}
\end{align}
where the ratios $\gamma_{3/5}$ are positive but \textit{not} bounded by unity. For our purpose, we focus on the regimes at a fixed and finite particle-particle nesting degree ($\gamma_5=1$), with the particle-hole nesting degree ranging from weak ($\gamma_3\sim 0.1$) to moderate ($\gamma_3\sim 10$).  
While we parametrize these ratios as two free parameters, physically these nesting degrees are controlled by the dispersions of different patches in a given moir\'e system, so they are tunable experimental parameters as a matter of principle.  

Finally, we can express the rest of the bare susceptibilities in terms of the above-mentioned $\Pi_{\rm ph}^{nn}(0)$, $\Pi_{\rm pp}^{n\bar{n}}(0)$, $\Pi_{\rm ph}^{n\bar{m}}(\textbf{Q}^-_{n\bar{m}})$, and $\Pi_{\rm pp}^{nm}(\textbf{Q}^-_{\bar{n}m})$ making use of the relations among patch center locations and the dispersions among patches 
\begin{align}
&\textbf{Q}'_n=\textbf{Q}^+_{nm}+\textbf{Q}^-_{n\bar{m}}\nonumber\\
&\epsilon^n_{\textbf{k}}\approx\epsilon^{\bar{n}}_{\textbf{k}+\textbf{Q}'_n}
\label{dispersionrelation}
\end{align}
where $n=1,2,3$ and $m$($\neq n$) are patch indices in the same valley and the latter is an equality up to second order in $k$. 
With these relations, we arrive at  
\begin{align}
&\Pi_{\rm ph}^{n\bar{n}}(\textbf{Q}'_n)=\gamma\Pi_{\rm ph}^{nn}(0),\\
&\Pi_{\rm ph}^{nm}(\textbf{Q}^+_{nm})=\gamma\Pi_{\rm ph}^{n\bar{m}}(\textbf{Q}^-_{n\bar{m}}),\\
&\Pi_{\rm pp}^{nn}(-\textbf{Q}'_n)=\gamma\Pi_{\rm pp}^{n\bar{n}}(0),\\
&\Pi_{\rm pp}^{n\bar{m}}(\textbf{Q}^+_{\bar{n}\bar{m}})=\gamma\Pi_{\rm pp}^{nm}(\textbf{Q}^-_{\bar{n}m}). 
\label{RelateBubbles}
\end{align}
Here we introduce a parameter $\gamma\in[0,1]$ to quantify the difference between 
the dispersions of opposite patches $\epsilon_{\textbf{q}}^n$ and $\epsilon_{\textbf{q}}^{\bar{n}}$, where $\textbf{q}$ is measured from the patch centers.  
How much $\gamma$ diviates from $1$ measures the magnitude of the cubic and higher-order corrections $O(k^3)$ to the band dispersions. 
When the van Hove singularity is purely quadratic, 
the two dispersions are identical and $\gamma=1$ [see Eq. \ref{dispersion}]. 
These relations among the bare susceptibilities are essential for deriving the RG equations of inter- and intra-patch interactions, and also for analyzing the competition among various instabilities.

\begin{figure}[t]
\includegraphics[width=8cm]{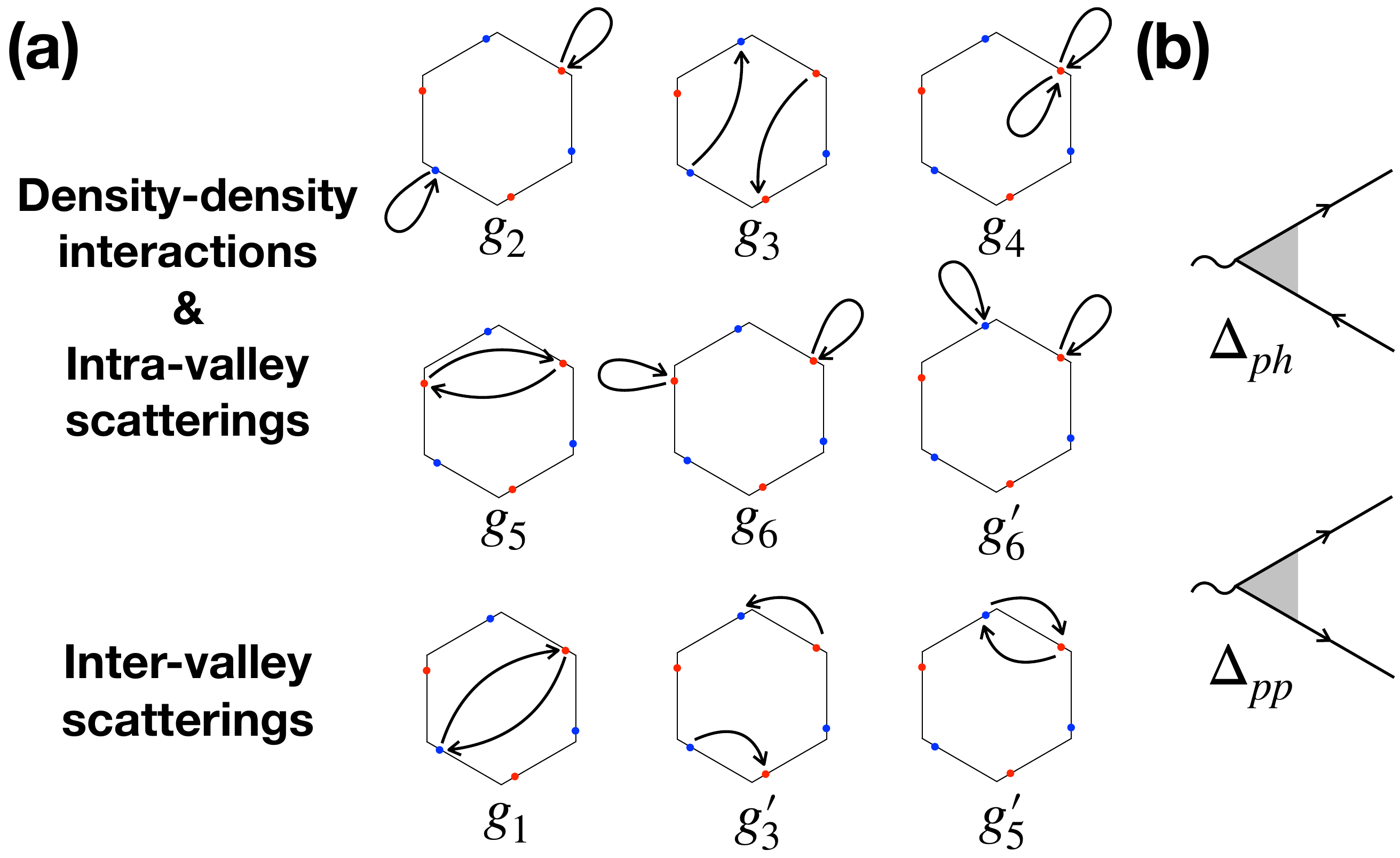}
\caption{
(a) Schematics for all the momentum-preserving inter- and intra-patch interactions. The hexagon, red dots, and blue dots represent the moir\'e Brillouin zone, the VH points from $K$ valley, and those from $-K$ valley, repectively. Diagrammatic expressions for (b) the test vertices in the particle-hole and particle-particle channels. 
}
\label{gi}
\end{figure}

\subsection{Inter- and intra-patch effective interactions}
With the bare susceptibilities in hand, we now explain the parquet RG approach we use to treat the considered moir\'e systems, which feature three VHS from each valley where the two valleys are related by time-reversal symmetry, as shown in Eq. \ref{epsilon1} and Fig. \ref{FS}(a)(b). 
When such systems are gated near these six VHS, the predominant contributions to the density of states come solely from the portion of FS near these VHS. We can therefore simplify the problem by making the ``patch approximation'', i.e. instead of the full BZ, considering only low-energy electrons living in patches centered at the van Hove points. Our approach is perturbative in the sense that the ultraviolet energy cutoff $\Lambda$ corresponding to the patch size $k_{\Lambda}$ is much smaller than the band width. 

Given these six patches, there are in total nine inequivalent inter- and intra-patch interactions allowed by the lattice symmetries and momentum conservation [see Fig. \ref{gi}(a) for schematics] 
\begin{align}
H_{\text{int}}=\frac{1}{2}\sum_{n=1}^3&\sum_{m\neq n}\sum_{ss'}
\tilde{g}_1\psi^{\dagger}_{\bar{n}s}\psi^{\dagger}_{ns'}\psi_{\bar{n}s'}\psi_{ns}
+\tilde{g}_2\psi^{\dagger}_{ns}\psi^{\dagger}_{\bar{n}s'}\psi_{\bar{n}s'}\psi_{ns}\nonumber\\
&+\tilde{g}_3\psi^{\dagger}_{ms}\psi^{\dagger}_{\bar{m}s'}\psi_{\bar{n}s'}\psi_{ns}
+\tilde{g}_3'\psi^{\dagger}_{\bar{m}s}\psi^{\dagger}_{ms'}\psi_{\bar{n}s'}\psi_{ns}\nonumber\\
&+\tilde{g}_4\psi^{\dagger}_{ns}\psi^{\dagger}_{ns'}\psi_{ns'}\psi_{ns}
+\tilde{g}_5\psi^{\dagger}_{ms}\psi^{\dagger}_{ns'}\psi_{ms'}\psi_{ns}\nonumber\\
&+\tilde{g}_5'\psi^{\dagger}_{\bar{m}s}\psi^{\dagger}_{ns'}\psi_{\bar{m}s'}\psi_{ns} 
+\tilde{g}_6\psi^{\dagger}_{ns}\psi^{\dagger}_{ms'}\psi_{ms'}\psi_{ns}\nonumber\\
&+\tilde{g}_6'\psi^{\dagger}_{ns}\psi^{\dagger}_{\bar{m}s'}\psi_{\bar{m}s'}\psi_{ns}, 
\end{align}
where $\psi_{ns}$ is the fermionic field for the electron on patch $n=1,2,3$ with spin $s=\uparrow,\downarrow$, patch $\bar{n}$ centers at the opposite momentum to patch $n$ (and thus from the other valley), and patch $m\neq n$ but belongs to the same valley as patch $n$.
Among these nine interactions, 
$\tilde{g}_4$, $\tilde{g}_2$, $\tilde{g}_6$, and $\tilde{g}_6'$ are density-density interactions, and the rest are scattering processes. More specifically, one has to consider density-density interactions for electrons within the same patch ($g_4$), between intra-valley patches ($g_6$), and between inter-valley patches ($g_2$ and $g_6'$).
As for the scattering processes, $\tilde{g}_3$ and $\tilde{g}_5$ are scatterings with intra-valley momentum transfer $\textbf{Q}^+$, whereas $\tilde{g}_1$, $\tilde{g}_3'$, and $\tilde{g}_5'$ are those with inter-valley momentum transfers $\textbf{Q}'$ and $\textbf{Q}^-$. In particular, $\tilde{g}_1$, $\tilde{g}_3$, and $\tilde{g}_3'$ are zero-momentum (BCS) pair scatterings, whereas $g_5'$ is that for finite-momentum pairs. 

The next step is to understand how these interactions $\tilde{g}_p$, $p=1,\cdots ,9$ among the patch electrons evolve as we decrease the energy towards the infrared limit. We show that such evolution is described by the following renormalization group (RG) equations up to the quadratic order 
\begin{widetext}
\begin{align}
&\frac{dg_1}{dy}=-2g_1g_2-2(N_p-1)g_3g'_3+2d_1(y)[g_1g_4+(N_p-1)g_5g'_5]+d_2(y)(2g_1g_2-N_fg_1^2), \nonumber\\
&\frac{dg_2}{dy}=-(g_1^2+g_2^2)-(N_p-1)(g_3^2+g_3'^2)+2d_1(y)[(1-N_f)g_2g_4+g_1g_4+(N_p-1)(g_5g'_6+g'_5g_6-N_fg_6g'_6)]+d_2(y)g_2^2,\nonumber\\
&\frac{dg_3}{dy}=-[2g_2g_3+2g_1g'_3+(N_p-2)(g_3^2+g_3'^2)]+2d_3(y)g_3g'_6+2\tilde{d}_3(y)(g_3g_6+g'_3g_5-N_fg_3g_5),\nonumber\\
&\frac{dg'_3}{dy}=-2[g_2g'_3+g_1g_3+(N_p-2)g_3g'_3]+2d_3(y)(g'_3g'_6+g_3g'_5-N_fg'_3g'_5)+2\tilde{d}_3(y)g'_3g_6,\nonumber\\
&\frac{dg_4}{dy}=-d_4(y)g_4^2+d_1(y)[g_1^2-N_fg_2^2+(3-N_f)g_4^2+(N_p-1)(g_5^2+g_5'^2+2g_5g_6+2g'_5g'_6)-(N_p-1)N_f(g_6^2+g_6'^2)+2g_1g_2],\nonumber\\
&\frac{dg_5}{dy}=-2d_5(y)g_5g_6+d_1(y)[2g_4g_5+2g_1g'_5+(N_p-2)(g_5^2+g_5'^2)]+\tilde{d}_3(y)[2g_5g_6+2g_3g'_3-N_f(g_5^2+g_3^2)],\nonumber\\
&\frac{dg'_5}{dy}=2d_1(y)[g_4g'_5+g_1g_5+(N_p-2)g_5g'_5]+d_3(y)[2(g'_5g'_6+g_3g'_3)-N_f(g_5'^2+g_3'^2)]-2\tilde{d}_5(y)g'_5g'_6,\nonumber\\
&\frac{dg_6}{dy}=2d_1(y)[(1-N_f)g_4g_6+g_4g_5+g_1g'_6+g_2g'_5-N_fg_2g'_6]
+d_1(y)(N_p-2)[2(g_5g_6+g'_5g'_6)-N_f(g_6^2+g_6'^2)]\nonumber\\
&~~~~~~+\tilde{d}_3(y)(g_3'^{2}+g_6^2)-d_5(y)(g_5^2+g_6^2),\nonumber\\
&\frac{dg'_6}{dy}=2d_1(y)[(1-N_f)g_4g'_6+g_4g'_5+g_1g_6+g_2g_5-N_fg_2g_6]+2d_1(y)(N_p-2)(g_5g'_6+g'_5g_6-N_fg_6g'_6)\nonumber\\
&~~~~~~+d_3(y)(g_3^2+g_6'^2)-\tilde{d}_5(y)(g_5'^2+g_6'^2),
\label{giRGeqn}
\end{align}
\end{widetext}
where $N_f=2$ and $N_p=3$ are the number of fermion flavor and number of patches per valley, respectively. Here, we define the RG running paramater to be $y\equiv \frac{1}{2}\rm{ln}^2$$(\frac{\Lambda}{E})\sim\Pi^{n\bar{n}}_{pp}(0)/\nu_0$, which is negatively related to the energy $E$, and $g_p\equiv\nu_0 \tilde{g}_p$ denotes the dimensionless interactions corresponding to interactions $\tilde{g}_p$ in Fig. \ref{gi}(a). 

In the above RG equations, we introduce the energy-dependent d factors $d_j(y)$, $j=1\cdots 5$, which capture the relative magnitudes between different bare susceptibilities and the RG running parameter $y$ as follows:
\begin{align}
&d_1(y)\equiv\frac{1}{\nu_0}\frac{d\Pi^{nn}_{ph}(0)}{dy},
~~~~~~~~~~~~d_2(y)\equiv\frac{1}{\nu_0}\frac{d\Pi^{n\bar{n}}_{ph}(\textbf{Q}'_{n})}{dy},\nonumber\\
&d_3(y)\equiv\frac{1}{\nu_0}\frac{d\Pi^{n\bar{m}}_{ph}(\textbf{Q}^-_{n\bar{m}})}{dy},
~~~~~~\tilde{d}_3(y)\equiv\frac{1}{\nu_0}\frac{d\Pi^{nm}_{ph}(\textbf{Q}^+_{nm})}{dy},\nonumber\\
&d_4(y)\equiv\frac{1}{\nu_0}\frac{d\Pi^{nn}_{pp}(-\textbf{Q}'_{n})}{dy},
~~~~~~~d_5(y)\equiv\frac{1}{\nu_0}\frac{d\Pi^{nm}_{pp}(\textbf{Q}^-_{\bar{n}m})}{dy},\nonumber\\
&\tilde{d}_5(y)\equiv\frac{1}{\nu_0}\frac{d\Pi^{n\bar{m}}_{pp}(\textbf{Q}^+_{\bar{n}\bar{m}})}{dy}. 
\end{align}
These relations characterize the key features of low-energy band structures relevant to the RG flows. 
For instance, $d_1(y)$ and $d_3(y)$ ($\tilde{d}_3(y)$) describe how the density of states and the intra-valley (inter-valley) particle-hole nesting evolve with $y$, respectively. 

These d factors $d_j(y)$ generally decrease as the energy decreases towards the FS ($E=0$), and are therefore decreasing functions in the RG parameter $y$. 
Asymptotically, in the ultraviolet limit $y\rightarrow 0$ (energy $E\rightarrow\Lambda$) these functions behave as $d_j(y)\sim 1$, whereas when approaching the infrared limit $y\rightarrow\infty$ (energy $E\rightarrow 0$) they behave as $d_1(y)\sim\frac{1}{\sqrt{2y}}$, $d_2(y)\sim\frac{\gamma}{\sqrt{2y}}$, $d_3(y)\sim\frac{\gamma_3}{\sqrt{2y}}$, $\tilde{d}_3(y)\sim\frac{\gamma\gamma_3}{\sqrt{2y}}$, $d_4(y)\sim\gamma$, $d_5(y)\sim\frac{\gamma_5}{\sqrt{2y}}$, and $\tilde{d}_5(y)\sim\frac{\gamma\gamma_5}{\sqrt{2y}}$. Based on the above asymptotic behavior, we model these functions $d_j(y)$ as follows:
\begin{align}
&d_1(y)\sim\frac{1}{\sqrt{1+2y}},
~~~~~~d_2(y)\sim\frac{\gamma}{\sqrt{\gamma^2+2y}},\nonumber\\
&d_3(y)=\frac{\gamma_3}{\sqrt{\gamma_3^2+2y}},
~~~\tilde{d}_3(y)=\frac{\gamma\gamma_3}{\sqrt{\gamma^2\gamma_3^2+2y}},\nonumber\\
&d_4(y)=\frac{1+\gamma y}{1+y},
~~~~~~~~~~~~~~~d_5(y) =\frac{\gamma_5}{\sqrt{\gamma_5^2+2y}},\nonumber\\
&\tilde{d}_5(y)=\frac{\gamma\gamma_5}{\sqrt{\gamma^2\gamma_5^2+2y}}. 
\label{dfactor}  
\end{align}
By plugging in Eq. \ref{dfactor} to Eq. \ref{giRGeqn} and numerically solving the RG differential equations for the intra- and inter-patch interactions, we find that relevant interactions $g_p(y)$ flow to the strong coupling limit and diverge as approaching some critical scale $y_c$. Since $y_c$ corresponds to the critical energy scale at which the perturbative approach breaks down, this energy scale can be associated with the critical temperature $T_c$ at which the instabilities destablize the FS.

Importantly, this critical scale $y_c$, which sets a cutoff for the RG flows, generally depends on both the low-energy band structures and the considered initial values $g_i(y=0)$, and is \textit{not always large} ($y_c\ll\infty$).  
In cases with relatively smaller $y_c$, the RG flows can depend strongly on  the contribution from \textit{both} ln- and ln$^2$-divergent susceptibilities in the RG equations since the diverging rates of ln and ln$^2$ are then comparable (such that $d_j(y)\lesssim 1$) for $y\leq y_c\ll\infty$. 
We therefore emphasize that it is necessary to keep all the terms associated with ln-divergent susceptibilities to obtain the correct RG flows and the dominant instabilities.  

To discern which interactions have higher divergence rates, we parameterize the interactions in the standard way as  
\begin{align}
g_p(y)=\frac{G_p}{y_c-y}.
\label{Gp}
\end{align}
In the following, we will study the dominant instabilities in terms of the effective interaction strengths $G_p$ at $y\rightarrow y_c$.

\subsection{Instabilities}
With the RG flows of the inter- and intra-patch interactions in hand, we are now ready to study the possibile instabilities in the system to identify the most dominant one. 
To this end, we first write down the test vertices for the instabilities in both particle-particle and particle-hole channels [see Fig. \ref{gi}(b)], then study the RG flows of these vertices in the infrared limit to see if the vertices are relevant or not.  

For the particle-particle instabilities, we consider uniform superconductivity (SC) and pair density waves (PDW) with test vertices
\begin{align}
&\Delta_{SC}^{n}\psi^{\dagger}_{\bar{n}s}\sigma^{i}_{ss'}\psi^{\dagger}_{ns'}~~,
~~~~~~\Delta_{PDW_{a}}^{n}\psi^{\dagger}_{ns}\sigma^{i}_{ss'}\psi^{\dagger}_{ns'}~~,\nonumber\\
&\Delta_{PDW_{b}}^{n}\psi^{\dagger}_{ns'}\sigma^{i}_{s's}\psi^{\dagger}_{ms}~~,
~~~\Delta_{PDW_{c}}^{n}\psi^{\dagger}_{ns}\sigma^{i}_{ss'}\psi^{\dagger}_{\bar{m}s'}, 
\end{align}
where $\sigma^i$ denotes the Pauli matrices in spin, and $i=0,x,y,z$. 
Here, $n$, $m\neq n$ label patches from the same valley, and subscripts $a$, $b$, $c$ in PDW indicate finite momenta pairs consisting of electrons from the same patch, different patches from the same valley, and patches from opposite valleys. 
As for the particle-hole channel, we consider both the magnetic and charge instabilities with zero and finite momentum transfers, namely the ferromagnetic instabilities (FM), the uniform charge orders (UC), and spin and charge density waves (SDW, CDW). 
The test vertices for magnetic instabilities have the following forms:  
\begin{align}
&\Delta_{FM}^{n}\psi^{\dagger}_{ns'}\sigma^j_{s's}\psi_{ns}~~,
~~~~~~\Delta_{SDW_a}^{n}\psi^{\dagger}_{ns'}\sigma^j_{s's}\psi_{\bar{n}s}~~,\nonumber\\
&\Delta_{SDW_b}^{n}\psi^{\dagger}_{ns'}\sigma^j_{s's}\psi_{ms}~~,
~~~\Delta_{SDW_c}^{n}\psi^{\dagger}_{ns'}\sigma^j_{s's}\psi_{\bar{m}s},
\end{align}
where $j=x,y,z$, and those for charge instabilities have the form   
\begin{align}
&\Delta_{C}^{n}\psi^{\dagger}_{ns}\sigma^0_{ss'}\psi_{ns'}~~,
~~~~~~\Delta_{CDW_a}^{n}\psi^{\dagger}_{ns'}\sigma^0_{s's}\psi_{\bar{n}s}~~,\nonumber\\
&\Delta_{CDW_b}^{n}\psi^{\dagger}_{ns'}\sigma^0_{s's}\psi_{ms}~~,
~~~\Delta_{CDW_c}^{n}\psi^{\dagger}_{ns'}\sigma^0_{s's}\psi_{\bar{m}s}. 
\end{align}
We consider density waves with both intra- and inter-valley momentum transfers $\textbf{Q}'$, $\textbf{Q}^{+}$, and $\textbf{Q}^{-}$, and label them with subscript $a$, $b$, and $c$ respectively.

We find that the vertex $\Delta_i$ for each instability $i$ renormalizes with the RG running parameter $y$ as 
$\frac{d\Delta_i}{dy}=-\beta_i\Delta_i$, where $\beta_i=d_{n(i)}\Gamma_i$ quantifies the tendency for this instability to dominate.   
Here $d_{n(i)}$, and $\Gamma_i$ are respectively the d factors [see Eq. \ref{dfactor}] associated with the relevant bare susceptibility, and the driving interaction for instability $i$. In particular, the driving interactions $\Gamma_i$ for different instabilities $i$ are given by different linear combinations of the inter- and intra-patch interactions $\{g_p\}$, and can be expressed in terms of $\{G_p\}$ defined in Eq. \ref{Gp}. 

This quantity $\beta_i$ quantifies the tendency for instability $i$ because 
it enters the renormalization of the susceptibility through $\frac{d\chi_i}{dy}=\tilde{d}_i|\Delta_i|^2$. Since the susceptibility evolve as $\chi_i\sim(y_c-y)^{\alpha_i}$ with $\alpha_i=2\beta_i+1$\cite{RG_FeSc_Chubukov_PRB,RG_Moire_Nandkishore}, 
it is clear that only instabilities with $\beta_i<-1/2$ are relevant, and  
the magnitude $|\beta_i|$ determines the diverging rate of the susceptibility. We therefore use $\beta_i$ as the measure to analyze the competition among instabilities. 
Among the relevant instabilities with $\beta_i<-1/2$, the instability $i$ with the most negative $\beta_i$ dominates in the infrared limit.   

We find the tendencies for the considered instabilities as follows. For particle-particle instabilities, 
\begin{align}
&\beta_{SC}^s=G_2+G_1+2(G_3+G'_3),~\beta^f_{SC}=G_2-G_1+2(G_3-G'_3)\nonumber\\
&\beta^d_{SC}=G_2+G_1-(G_3+G'_3),~~\beta^p_{SC}=G_2-G_1-(G_3-G'_3)\nonumber\\
&\beta_{PDW_a}=d_4(y_c)G_4,~\beta_{PDW_b}^{\pm}=d_5(y_c)(G_6\pm G_5),\nonumber\\
&\beta_{PDW_c}^{\pm}=\tilde{d}_5(y_c)(G'_6\pm G'_5),
\label{betaSC}
\end{align} 
where the superscripts indicate different pairing symmetries. In particular, the uniform superconductivity 
$\beta_{SC}^{s/d}$, the intra-patch PDW $\beta_{PDW_a}$, and the intra/inter-valley PDW  $\beta_{PDW_{b/c}}^{+}$ are associated with spin-singlet pairing. In contrast, the uniform superconductivity $\beta_{SC}^{f/p}$, and the intra/inter-valley PDW with sign changes in pairing potentials $\beta_{PDW_{b/c}}^{-}$ are associated with spin-triplet pairing. 

For the magnetic instabilities, we find 
\begin{align}
&\beta_{FM}^s=-d_1(y_c)[G_1+G_4+2(G_5+G'_5)],\nonumber\\
&\beta^f_{FM}=-d_1(y_c)[-G_1+G_4+2(G_5-G'_5)],\nonumber\\
&\beta^d_{FM}=-d_1(y_c)[G_1+G_4-(G_5+G'_5)],\nonumber\\
&\beta^p_{FM}=-d_1(y_c)[-G_1+G_4-(G_5-G'_5)],\nonumber\\
&\beta_{SDW_a}=-d_2(y_c)G_2,~\beta_{SDW_b}^{\pm}=-\tilde{d}_3(y_c)(G_6\pm G'_3),\nonumber\\
&\beta_{SDW_c}^{\pm}=-d_3(y_c)(G'_6\pm G_3), 
\label{betaS}
\end{align}
and for the charge instabilities, we find
\begin{align}
&\beta_{UC}^s=-d_1(y_c)[G_1-2G_2-G_4+2(G_5-2G_6+G'_5-2G'_6)],\nonumber\\
&\beta^f_{UC}=-d_1(y_c)[-G_1+2G_2-G_4+2(G_5-2G_6-G'_5+2G'_6)],\nonumber\\
&\beta^d_{UC}=-d_1(y_c)[G_1-2G_2-G_4-(G_5-2G_6+G'_5-2G'_6)],\nonumber\\
&\beta^p_{UC}=-d_1(y_c)[-G_1+2G_2-G_4-(G_5-2G_6-G'_5+2G'_6)],\nonumber\\
&\beta_{CDW_a}=-d_2(y_c)(G_2-2G_1),\nonumber\\
&\beta_{CDW_b}^{\pm}=-\tilde{d}_3(y_c)[G_6-2G_5\pm (G'_3-2G_3)],\nonumber\\
&\beta_{CDW_c}^{\pm}=-d_3(y_c)[G'_6-2G'_5\pm (G_3-2G'_3)].
\label{betaC}
\end{align}
Here, we have considered FM and UC instabilities with different form factors labeled by their superscripts $s/f/d/p$, and the superscripts $\pm$ for intra- and inter-valley density waves indicate whether the order parameters exhibit a sign change or not across patches.   

A few remarks about the instabilities listed above: First, the FM instabilities with $s$ and $f$ form factors respectively corresond to the \textit{spin-polarized ferromagnetic state} and  \textit{valley antiferromagnetic state}, where the latter has opposite spin polarizations in opposite valleys. The UC instability with $f$ form factor corresonds to a \textit{valley-polarized state}. 
Moreover, FM and UC instabilities with $p$ or $d$ form factors all break the threefold rotational symmetry and are thus associated with various \textit{nematic orders}.
Finally, the $d$ factors for the magnetic and charge instabilities carry a minus sign due to the fact that superconductivity and particle-hole instabilities are driven by attraction and repulsion, respectively. 

\begin{figure}[t]
\includegraphics[width=6cm]{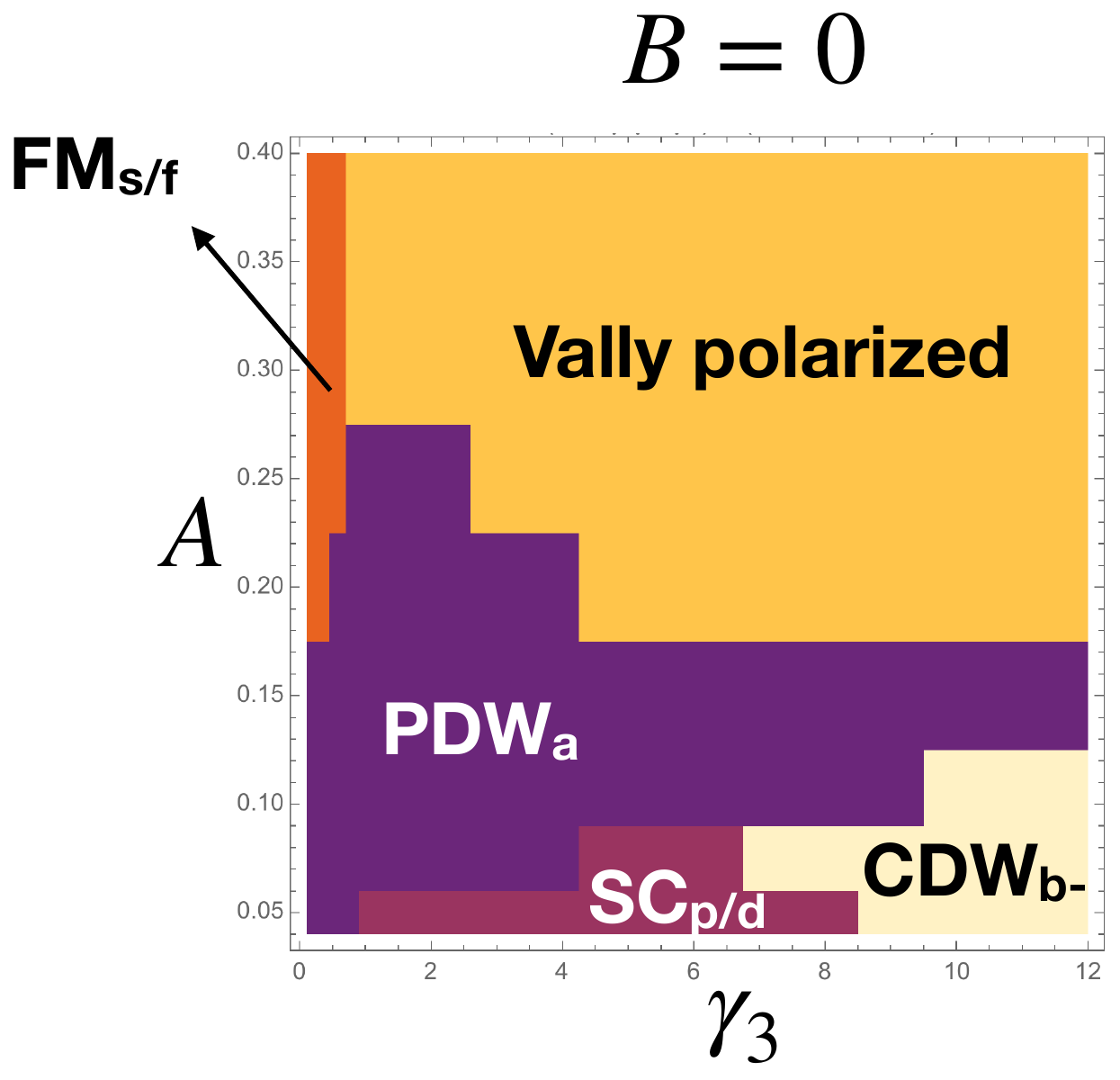}
\caption{
The phase diagram in the absence of inter-valley scattering ($B=0$).   
}
\label{PD1}
\end{figure}
\section{phase diagram}\label{sec:PD}
\subsection{Key parameters}
In this section, we present and discuss the dominant instabilities we find using the perturbative RG approach in the patch model we consider [see Fig. \ref{FS}(d)] when varying two important parameters. 
The first is the nesting degree in the particle-hole channel, which is parametrized by $\gamma_3$ in Eq.~\ref{dfactor}. 
Importantly, $\gamma_3$ can be larger than $1$ since it measures 
the nesting degree in reference to the magnitude of the ln-divergent density of states instead of the ln$^2$-divergent Cooper susceptibility.   
We choose to parametrize the nesting degree this way 
because we want to zoom in and focus on the regimes with nesting degree ranging from weak to moderate, but far from perfect.

The second key parameter is the interaction strength in the ultraviolet limit of the patch model, i.e. the initial values $g_i(y=0)$ we insert to the RG flows of the intra- and inter-patch interactions $g_i(y)$. 
For simplicity, we use only two variables to parametrize these initial values and assume all intra-valley scatterings $g_i(0)=A$ with $i=2,3,4,5,6,6'$, and all inter-valley scatterings $g_i(0)=B$, $i=1,3',5'$. In particular, we focus on the cases where both $A$ and $B$ are repulsive, and show results for two representative cases $B=0$ and $B=A/4$ given that the intra-valley scatterings are expected to be stronger than the inter-valley ones. 

The interaction strength $A$ is an important parameter because 
while the ln$^2$-divergent instabilities, such as uniform SC, always dominate in the weak-coupling limit, the ln-divergent instabilities, such as ferromagnetic instabilities, can become competitive and even dominant when $A$ is moderate. 
This is because, as the interaction strength $A$ increases from weak to moderate, 
the increasing critical temperature will impose an earlier cutoff to the RG flows of the interactions $g_i$ as well as the energy-dependent $d$ factors (see Eq. \ref{dfactor}). Given that the slow-growing ln$^2$ function may not be significantly larger than a ln function depending on how low the cutoff scale is, the tendencies $\beta_j$ for ln- and ln$^2$-divergent instabilies $j$ can be comparable, and the balance is essentially tilted by their respective driving interactions $\Gamma_j$. 
Consequently, in the moderate coupling regimes (which are still ``weak'' compared to band width), both particle-hole and particle-particle instabilities stay competitive and must both be taken into consideration.\\

\subsection{In the absence of inter-valley scattering}
We first study the dominant instabilities in the absence of inter-valley scattering, i.e. $B=0$. In this limit, each valley preserves its own SU(2) spin rotational symmetry, and therefore the system has an enlarged SU(2)$\times$SU(2) symmetry. Consequently, instabilities that can be transformed into each other by valley-dependent spin rotations become energetically degenerate\cite{Thy_DBLG_SCIrrep_Samajdar} and share the same $\beta_j$ [see Eq. \ref{betaSC}-\ref{betaC}]. Important instabilities that become degenerate are spin-singlet $d$-wave and triplet $p$-wave SC, FM with $s$-wave and $f$-wave form factors, and inter-valley spin and charge density waves.  

The phase diagram in the absence of inter-valley scattering under this SU(2)$\times$SU(2) symmetry is shown in Fig. \ref{PD1} in terms of the coupling strength $A$ and particle-hole nesting degree $\gamma_3$.  
In the weak coupling limit (small $A$), the dominant instabilities are the ln$^2$-divergent instabilities, namely the uniform SC and the PDW formed by two electrons on the same patch (PDW$_a$). The latter stays competitive with uniform SC because we assume the dispersions within opposite patches $n$ and $\bar{n}$, which come from opposite valleys, to be nearly degenerate.  
Such an assumption clearly holds when higher-order terms above $O(k^3)$ in the dispersions are negligible [see Eq. \ref{dispersion}]. 
In Fig. \ref{PD1}, we consider the limit where opposite patches $n$ and $\bar{n}$ have degenerate dispersions, which is mathematically described by setting $\gamma=1$. The d factor $d_4(y)$ for PDW$_a$ therefore becomes energy-independent and stays $1$, just as for the uniform SC. The competition between uniform SC and PDW$_a$ is thus solely controlled by their driving interactions $\Gamma_j$, which are determined by the RG flows of the intra- and inter-patch interactions.

\begin{figure}[t]
\includegraphics[width=8cm]{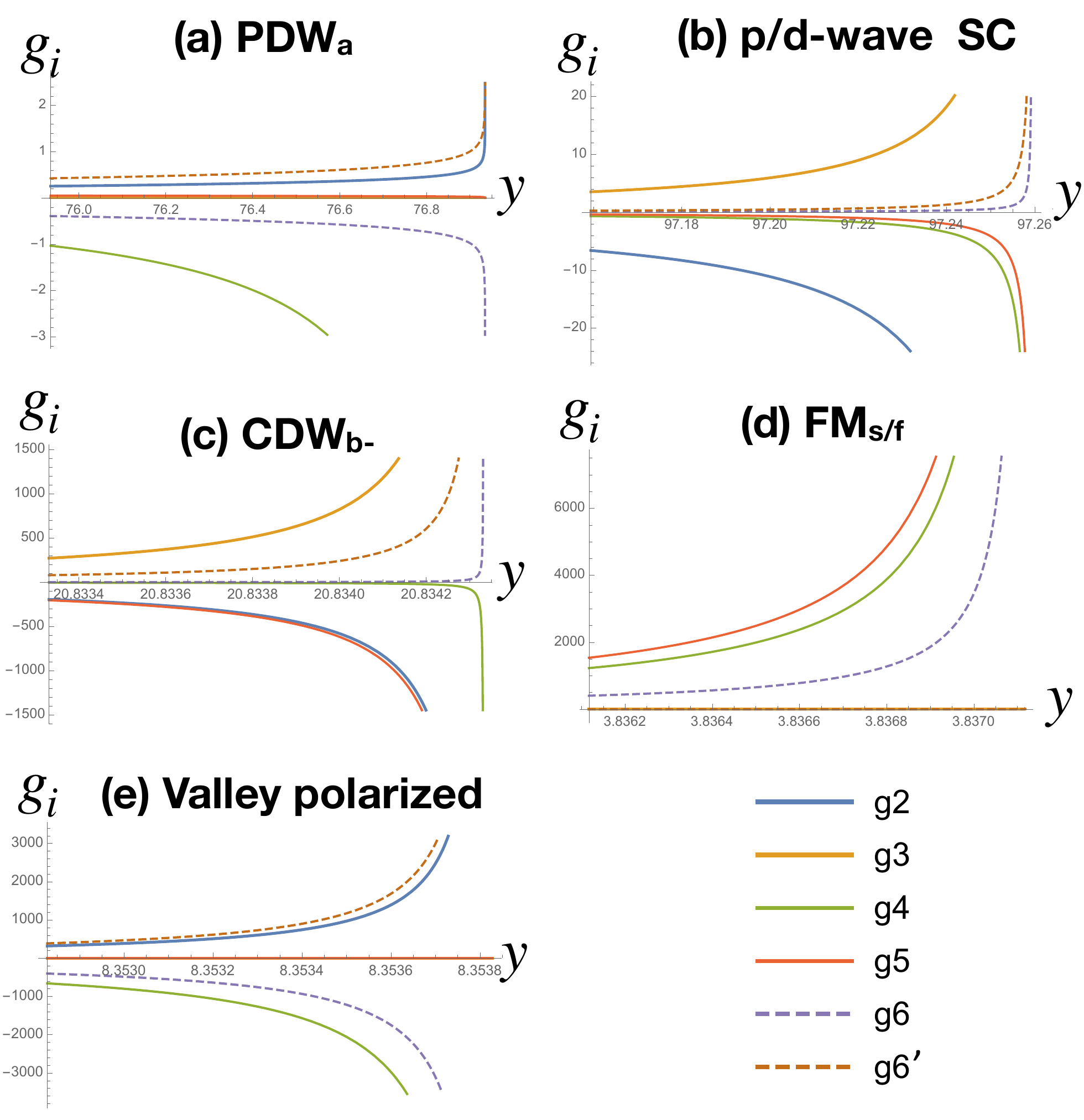}
\caption{Representative RG flows of intra- and inter-patch interactions in the absence of inter-valley scattering ($B=0$) in the regimes where the dominant instabilities are (a) intra-patch PDW, (b) $p/d$-wave uniform SC, (c) intra-valley CDW, (d) ferromagnetism with the $s/f$-wave form factor, and (e) valley-polarized uniform charge order, respectively.   
}
\label{gi1}
\end{figure}

In the weak nesting limit (small $\gamma_3^{(')}$), since the intra-patch density-density interaction $g_4$ is the only interaction whose RG equation is not directly impacted by the smallness of the nesting contribution, $g_4$ becomes the dominant relevant interaction [see Fig. \ref{gi1}(a)].  Together with the fact that in the weak-coupling limit the RG flows are dominated by the contributions in the Cooper channel, $g_4$ is attractive.
This intra-patch density-density attraction $g_4$ therefore is responsible for the dominant PDW$_a$. 

As the nesting degree increases, the $p$/$d$-wave uniform SC takes over instead of the $s$/$f$-wave one. This can be understood as follows. The $p$/$d$-wave SC differs from the $s$/$f$-wave ones in that the pairing potential change signs within a single valley for the former case  while that for the latter does not. Therefore, the balance between the $p$/$d$- and $s$/$f$-wave pairings is controlled by the intra-valley  scattering of BCS pairs $g_3$, and a potential with sign change is energetically favored by a repulsive $g_3>0$. 
In fact, $g_3$ is the interaction that receives most nesting-related repulsive contribution to its RG flow [see Fig. \ref{gi1}(b)]. The $p$/$d$-wave uniform SC thus dominates
in the moderate nesting regime. 
It is worth emphasizing that such a uniform SC is two-fold degenerate, and is expected to be spontaneously time-reversal broken due to energetic reasons. 
This chiral $p$/$d$-wave SC is known to be topological. 

As the nesting degree becomes even stronger (moderate $\gamma_3^{(')}$), not only the d factors for the density waves ($d_3^{(')}$) approach $1$, the intra-patch exchange interaction $g_5$ also becomes a strongly relevant attraction [see Fig. \ref{gi1}(c)]. 
The two factors together overcome the tendency of uniform SC and boost the intra-valley charge density wave CDW$_{b-}$ with sign change in the order parameter. This charge density wave does not have a degenerate spin density wave partner since it comprises two electrons from the same valley, which transform together under the single-valley spin rotation. 
Note that this charge density wave CDW$_{b-}$ we find correspond to the CDW$^+_-$ phase and the imaginary CDW phase found in Ref. \onlinecite{RG_Moire_Nandkishore} in strongly nested regimes. 

We now turn to the moderate coupling regime (moderate $A$). 
As the interaction strength $A$ increases, the ln-divergent instabilities begin to join the competition with the pairing instabilities since the enhanced critical temperature sets an earlier cutoff to the RG flows. At a low enough cutoff, ln- and ln$^2$-divergent instabilities may have comparable $\beta_j$ because the difference between their d factors becomes comparable to that between their driving interactions. The particle-hole instabilities therefore can now dominate over pairing instabilities even in regimes with a weak or moderate nesting degree. 

In the weak nesting limit, the intra-patch density-density interaction $g_4$ is still the dominant relevant interaction, similar to the regime where PDW$_a$ dominates.   
However, as $A$ increases, the density-of-states-related contributions (terms with $d_1(y)$ and $d_2(y)$) to the RG flows become non-negligible and flip $g_4$ from an attraction to a repulsion [see Fig. \ref{gi1}(d)]. This attractive $g_4$ drives only the ferromagnetic instabilities, and is therefore responsible for the dominance of FM in the weak nesting limit. 
Additionally, the intra-valley exchange interaction $g_5$ also receives sizable repulsive corrections from the density-of-states-related contributions [see Fig. \ref{gi1}(d)]. This repulsive $g_5$ further selects the ferromagnetic states whose order parameter has no sign change within a single valley (FM$_{s/f}$). 

As we further increase the nesting degree in the moderate coupling regime, we find that a \textit{valley-polarized} state, which corresponds to the uniform charge order with an $f$-wave form factor, dominates over the spin-polarized ferromagnetic states.  
A uniform charge order is favored over a uniform spin order because as the nesting degree increases, 
the test vertex $\Delta_{C}^f$ of the former receives an enhancing contribution 
from the inter-valley density-density interactions ($g_2$ and $g_6'$) [see Fig. \ref{gi1}(e)] that is non-vanishing only in charge channel. 
In fact, besides the spin-polarized ferromagnetism\cite{Exp_DBLG_Kim,Exp_SC_DBLG_Zhang19,Exp_DBLG_Pablo19,Exp_VPCI_DBLG_Tutuc19}, insulating phases with valley polarization or spin and valley polarization were also suggested at commensurate filling factors in experiments on TDBG\cite{Exp_DBLG_Pablo19,Exp_VPCI_DBLG_Tutuc19}. The phases we find here are generically metallic, but can become insulating when the carrier density corresponds to a commensurate filling factor\cite{wu2019Ferro,Wu2020Coll} because of the existence of spin (or valley) gaps. 

\begin{figure}[t]
\includegraphics[width=5.5cm]{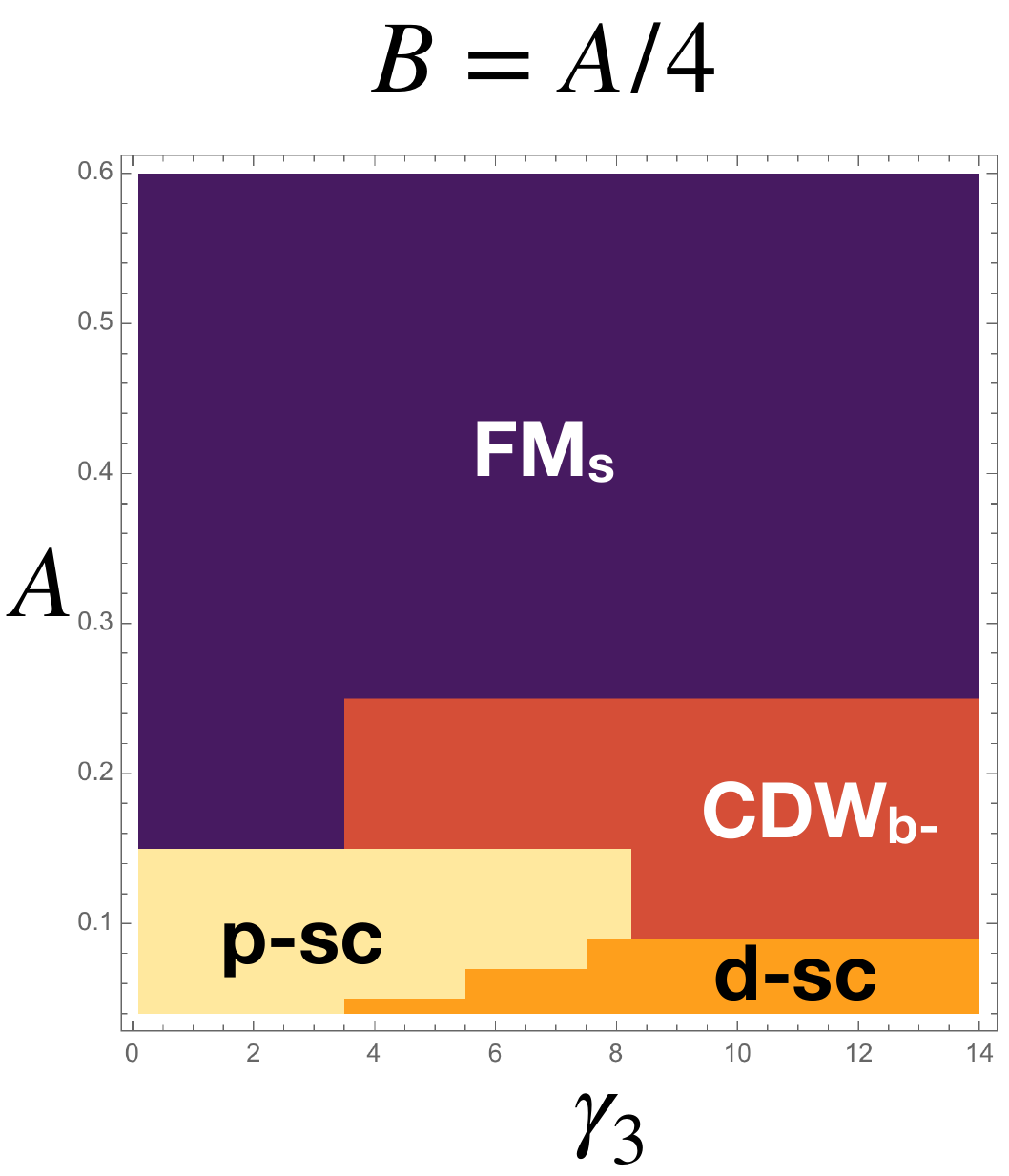}
\caption{The phase diagram in the presence of inter-valley scattering ($B=A/4$).   
}
\label{PD2}
\end{figure}
\begin{figure}[t]
\includegraphics[width=8cm]{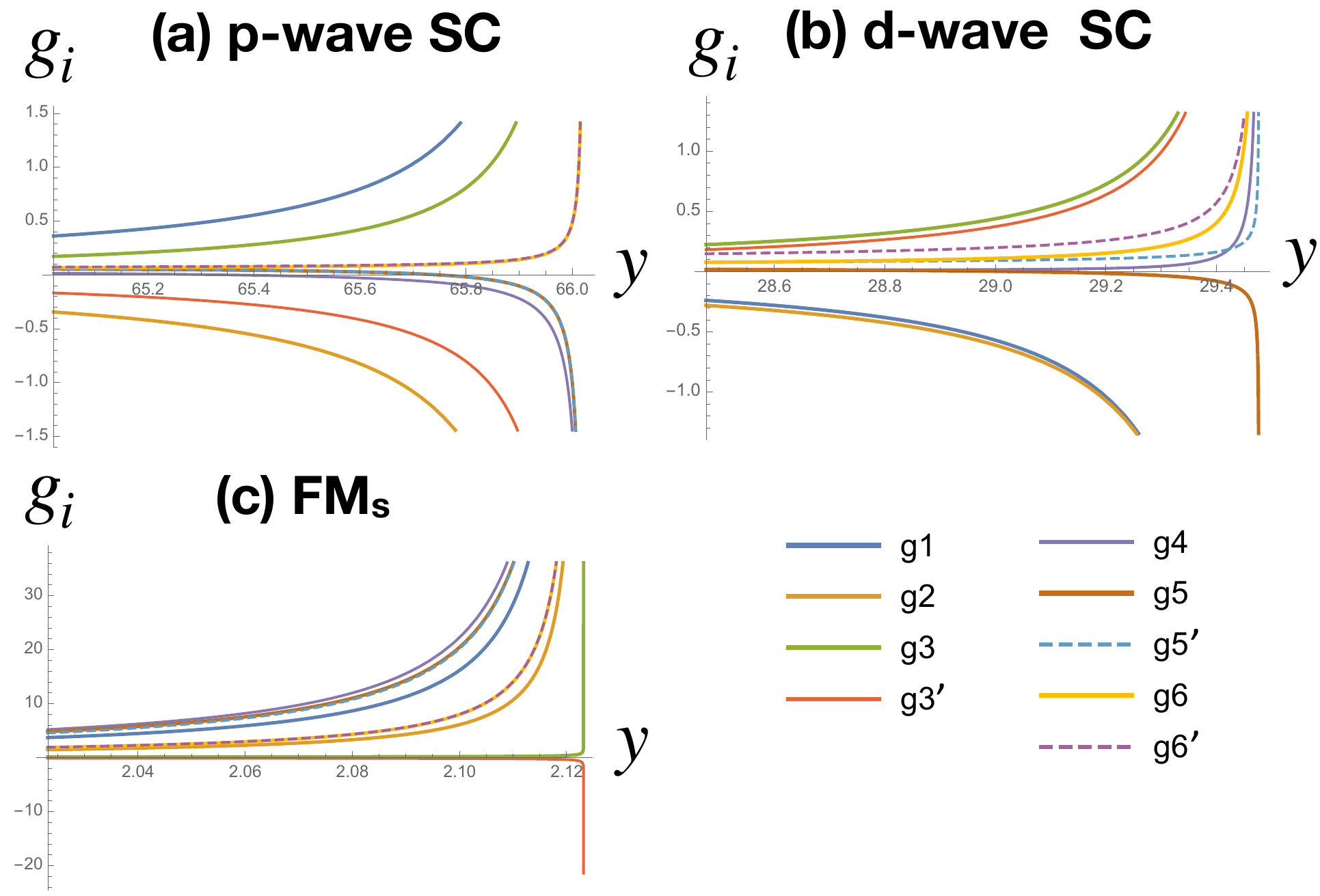}
\caption{Representative RG flows of intra- and inter-patch interactions in the presence of inter-valley scattering ($B=A/4$) in the regimes where the dominant instabilities are (a) $p$-wave uniform SC, (b) $d$-wave uniform SC, and (c) ferromagnetism, respectively.}
\label{gi2}
\end{figure}
\subsection{Inter-valley scattering}
We now study how dominant instabilities are affected by the presence of inter-valley scatterings, which include 
exchange processes with inter-valley momentum transfers $\textbf{Q}'$ and $\textbf{Q}^-$ between BCS pairs ($g_1$, $g_3'$) and finite-momentum pairs ($g_5'$) [see Fig. \ref{gi}(a)]. 
There are two main effects of having finite but small repulsive initial values for the RG flows of these interactions (i.e. $0<B<A$).  
First, these interactions can now break the SU(2)$\times$SU(2) symmetry if they become relevant, splitting the degeneracy between the following instabilities:  
the spin-singlet and triplet uniform SC, ferromagnetic instabilities with different form factors, and the inter-valley density waves in the spin and charge channels. 
Second, the intra-valley scatterings may also receive sizable second-order corrections from these inter-valley scatterings and become less or more relevant, or even change signs (because internal loops may contain inter-valley scattering terms). The landscape of the driving interactions for different instabilities can therefore undergo qualitative changes, and significantly impact the phase diagram.    

We present the resulting phase diagram in Fig. \ref{PD2}. 
One significant change in this phase diagram compared to Fig. \ref{PD1} is that the uniform SC now 
becomes the only instability in the weak-coupling limit (small $A$). 
This is because superconductivity with zero pair momentum is driven by scattering interactions of pairs on opposite patches $n$ and $\bar{n}$, which include the BCS pair exchanges with momentum transfer $\textbf{Q}'$ ($g_1$) and $\textbf{Q}^{\pm}$ ($g_3$, $g_3'$), and the density-density interaction ($g_2$) [see Eq. \ref{betaSC}]. 
Since these interactions are heavily coupled to each other, they are all enhanced and become the most relevant interactions when finite inter-valley scatterings $g_1$ and $g_3'$ are introduced. 
The tendency of uniform SC is therefore enhanced by the presence of inter-valley scatterings.  

Importantly, the inter-valley scatterings $g_1$ and $g_3'$ can further split the degeneracy between spin-singlet and triplet SC, and which pairing symmetry is most dominant is determined by whether these scatterings are repulsive or attractive. 
In particular, spin-singlet (spin-triplet) SC, which has pairing potentials with opposite (same) signs on opposite patches $n$ and $\bar{n}$, is energetically favored by a repulsive (attractive) pair exchange between $n$ and $\bar{n}$. 
Then an attractive (repulsive) inter-valley BCS pair scattering $g_3'$ can further promote $p$-wave ($d$-wave) pairing due to the pair potential sign changes among $n$ and other opposite-valley patches $\bar{m}$, $m\neq n$. 
In the weak nesting limit where $\gamma_3$ is small, 
$g_1$ becomes a relevant repulsion [see Fig. \ref{gi2}(a)] due to its negative coupling to other pair scatterings 
in the Cooper channel. 
This further leads to an attractive $g_3'$ [see Fig. \ref{gi2}(a)] due to its negative coupling to $g_1$. 
The $p$-wave SC thus dominates over the $d$-wave one. 

As the nesting degree increases, the corresponding inter-patch density-density interaction $g_6'$ receives repulsive enhancement. The inter-valley BCS pair scattering $g_3'$ then experiences a sign change and becomes a relevant repulsion through its coupling to $g_6'$ [see Fig. \ref{gi2}(b)]. Again due to its negative coupling to $g_3'$, the pair exchange $g_1$ also undergoes a sign change and becomes a relevant attraction [see Fig. \ref{gi2}(b)]. 
$d$-wave SC thus dominates over $p$-wave in the moderate nesting regime. 
Importantly, both $p$- and $d$-wave SC are doubly degenerate since they are both in two-dimensional representations of the point group $C_3$. Based on energetics, we therefore expect the uniform pairing in the weak coupling regime to be topological chiral $p$- and $d$-wave SC. 

Another significant change in the phase diagram due to the presence of the inter-valley scatterings is that the spin-polarized ferromagnetic instability now 
dominates the entire moderate-coupling regime [see Fig. \ref{PD1} and \ref{PD2}]. 
The key reason that tilts the balance between the spin-polarized state and the valley-polarized charge order is the intra-patch density-density interaction $g_4$: 
while the former is driven by a repulsive $g_4$, the latter is driven by an attractive $g_4$. This is due to an attractive density-density correction from $g_4$ to the RG flows of the test vertex $\Delta_C$ that is non-vanishing only in the charge channel (see the corresponding tendencies in Eq. \ref{betaS}-\ref{betaC}). 
In the absence of inter-valley scatterings ($B=0$), $g_4$ is a relevant attraction [see Fig. \ref{gi1}(e)] mainly due to the correction in the Cooper channel its RG flow receives. 
In the presence of inter-valley scatterings ($B\neq 0$), however, $g_4$ receives extra repulsive contributions from the inter-valley exchanges $g_1$ and $g_5$ that are related to density of states [see Eq. \ref{giRGeqn}]. 
These repulsive corrections become most significant and turn $g_4$ into a relevant repulsion [see Fig. \ref{gi2}(c)] in the moderate-coupling regime, where the critical temperature becomes significant enough such that the ln-divergent susceptibility, such as the density of states, becomes parametrically non-negligible. 
Therefore in the presence of intervalley scatterings, the valley-polarized charge order is suppressed and the spin-polarized ferromagnetic instability dominates over the entire moderate-coupling regime. 

\section{Summary and discussion}\label{sec:summary}
In summary, we conduct a perturbative RG group analysis on a “hot-spot”-type patch model associated with van Hove singularities for TDBG to investigate the dominant instabilities under two varying parameters: the repulsive interaction strengths relative to the band width and the Fermi surface nesting degrees. 
In particular, we focus on a range of interaction strength from infinitesimal to weak but finite and a nesting degree from weak to moderate, motivated by the observed small gap size \cite{Exp_DBLG_Kim,Exp_DBLG_Pablo19,Exp_SC_DBLG_Zhang19,Exp_VPCI_DBLG_Tutuc19} and the fact that such nesting degree is allowed under the lattice symmetry\cite{RG_Moire_Nandkishore}. 
The contribution from electrons away from the van-Hove patches is expected to increase for systems with intermediate interaction strength\cite{edgefermion_Nandkishore}. Extending our study to include these contributions, such as considering electrons living on the Fermi surface edges, is left as an interesting future direction. 

In the absence of inter-valley scatterings, we find that $d/p$-wave topological  superconductivity is likely favored for infinitesimal interactions, whereas an exotic modulated intra-patch paired state and spin- or valley-polarized metallic phases gain dominance as the interaction strength increases. 
When small inter-valley scatterings are turned on, we find that degeneracies between various phases are broken as expected. Consequently, $p$- and $d$-wave topological superconductivity dominate the weak and moderate nesting regimes for infinitesimal interactions, whereas the spin-polarized phase is predominant over the regime from weak to moderate interactions. 

The two parameters we explore, namely the relative interaction strength and the nesting degree, are both experimentally tunable. Specifically, the former can be experimentally tuned by the angle-dependent band width\cite{Thy_DBLG_ncomm}, whereas the latter can be controlled by the displacement field \cite{wu2019Ferro}, as shown in Fig. \ref{FS}(a)(b). 
Although a quantitative comparison between experimental and theoretical parameters is difficult (since experimental details vary quite a bit from sample to sample, indicating that the experimental parameters are not yet unique), we expect that samples with larger angles and smaller displacement field could more easily host the superconducting phases we predict. It is possible that the recently observed superconductivity in TDBG may very well be our predicted SC phase, but much more work is necessary to validate this idea since electron-phonon interaction may also produce superconductivity in TDBG\cite{Thy_phonon_DBLG_Xiao}.

Moreover, we expect the spin- or valley-polarized instabilities to be in general metallic, although they can become insulating at commensurate fillings. These interesting metallic phases in the TDBG or similar moir\'e systems could be potentially useful for application purposes in spin- and valley-tronics. 
We therefore urge experimental efforts for detecting spin or valley polarization in the observed metallic phases in TDBG\cite{Exp_SC_DBLG_Zhang19,Exp_DBLG_Kim,Exp_DBLG_Pablo19,Exp_DBLG_Cmetallic} by measurements such as ferromagnetic resonance, anomalous Hall effect, and Kerr rotation, while tuning the twist angle, carrier density, and displacement field. 

Finally, since our patch model and the perturbative RG approach depend only on the properties of the van Hove singularities near the Fermi level in TDBG, we expect our results to be general for systems with a similar van Hove fermiology. 
Specifically, such van Hove fermiology contains three van Hove points per species (valley in the TDBG case) that are related by three-fold rotation, and two species related by time-reversal symmetry. 
For instance, we expect our findings to hold qualitatively in the presence of lattice relaxation effects, which can be sizable in twisted bilayer systems. 
This is because the lattice relaxation generally preserves the three-fold rotational symmetry\cite{latticerelax}, and therefore preserves the van Hove fermiology we study. 

In our RG studies, the key parameters we explore are the nesting degree in particle-hole channel, the initial intra-valley interaction strength, and the inter-valley interactions. 
There are in fact a few other parameters that would also affect the RG results. Besides the number of patches, species, and the flavor of fermions (2 for the TDBG case since electrons are spin-1/2), the nesting in the particle-particle channel and the corrections beyond the quadratic order to the dispersions are also interesting parameters for future exploration.\\ 

\emph{Acknowledgment}---YTH thanks Yu-Ping Lin for very helpful discussions. We acknowledge support by the Laboratory for Physical Sciences.

%

\end{document}